

\magnification=\magstep1
\voffset=1.00truein
\settabs 18 \columns
\hoffset=1.00truein
\baselineskip=17 pt
\ifnum\pageno=1
\topinsert \vskip 0.50 in
\endinsert
\fi

\def\b{\bigskip}
\def\bb{\bigskip\bigskip}

\def\mybox{\sqcap\kern-.66em\sqcup\kern.66em}

\def\ce{\centerline}
\def\ve{\vfill\eject}

\def\e e{$e^+ e^-$ }


\ce {\bf THE GREENING OF QUANTUM FIELD THEORY}
\ce {\bf GEORGE AND I\footnote{*}{\rm{Lecture of July 14, 1993, at
Nottingham.}}}
\b
\ce {Julian Schwinger}
\ce {\it University of California, Los Angeles, CA 90024-1547}

\bb

The young theoretical physicists of a generation or two earlier subscribed to
the belief that: If you haven't done something important by age 30, you never
will.  Obviously, they were unfamiliar with the history of George Green, the
miller of Nottingham.

Born, as we all know, exactly two centuries ago, he received, from the age 8,
only a few terms of formal education.  Thus, he was self-educated in
mathematics and physics, when in 1828, at age 35, he published, by
subscription, his first and most important work: {\it An Essay on
the Applications of Mathematical Analysis to the Theory of Electricity and
Magnetism}.  The Essay was dedicated to a noble patron of the ``Sciences and
Literature", the Duke of Newcastle.  Green sent his own copy to the Duke.  I do
not know if it was acknowledged.  Indeed, as Albert Einstein is cited as
effectively saying, during his 1930 visit to Nottingham, Green, in writing the
Essay, was years ahead of his time.

There are those who cannot accept that someone, of modest social status and
limited formal education, could produce formidable feats of intellect.  There
is the familiar example of William Shakespeare of Stratford on Avon.  It took
almost a century and a half to surface, and yet another century to strongly
promote, the idea that Will of Stratford could not possibly be the source of
the plays and the sonnets which had to have been written by Francis Bacon.  Or
was it the earl of Rutland?  Or perhaps it was William, the sixth earl of
Derby?  The most recent pretender is Edward deVir, Seventeenth earl of Oxford,
notwithstanding the fact that he had been dead for 12 years when Will was put
to rest.

I have always been surprised that no one has suggested an analogous conspiracy
to explain the remarkable mathematical feats of the Miller of Nottingham.  So I
invented one.

Descended from one of the lines of the earl of Nottingham was the branch of the
earls of Effindham, which was separated from the Howards in 1731.  The fourth
holder of the title died in 1816, with apparently no claimant.  In that year,
George Green, age 23, could well have reached the maturity that led, 12 years
later, to the publication of the Essay.  And what of the remarkable fact that,
in the same year that the earldom was revived, 1837, George Green graduated
fourth wrangler at Cambridge University?

The conspiracy at which I hint darkly is one in which I believe quite as much
as I think Edward deVir is the real Shakespeare.

I consider myself to be largely self-educated.  A major source of information
came from my family's possession of the Encyclopedia Brittanica Eleventh
Edition.  I recently became curious to know what I might have, and probably
did, learn about George Green, some 65 years before.

There is no article detailing the life of George Green.  There are, however, 4
brief references that indicate the wide range of Green's interests.

First, in the article Electricity, as a footnote to the description of Lord
Kelvin's work, is this:

In this connexion the work of George Green (1793-1841) must not be forgotten.
Green's {\it Essay on the application of mathematical analysis to the theories
of electricity and magnetism}, published in 1828, contains the first exposition
of the theory of potential.  An important theorem contained in it is known as
Green's theorem, and is of great value.

It was, of course, Lord Kelvin, or rather William Thomson, who rescued Green's
work from total obscurity.

Then, in the article Hydromechanics, after several applications of Green's
transformation, which is to say, the theorem, there appears, under the heading
{\it The Motion of a Solid through a Liquid}:

The ellipsoid was the shape first worked out, by George Green, in his {\it
Research on the vibration of a pendulum in a fluid medium (1833)}.

On to the article Light under the heading {\it Mechanical Models of the
Electromagnetic Medium}.  After some negative remarks about Fresnel, one reads:

Thus, George Green, who was the first to apply the theory of elasticity in an
unobjectional manner ...

This is the content of {\it On the Laws of Reflexion and Refraction of Light
(1837)}.

Finally, the paper {\it On the Propagation of Light in Crystallized Media
(1839)} appears in the Brittanica article Wave as follows:

The theory of waves diverging from a center in an unlimited crystaline medium
has been investigated with a view to optical theory by G. Green.

The word ``propagation" is a signal to us that, in little more than 10 years,
George Green had significantly widened his physical framework.  From the static
three-dimensional Green function that appears in potential theory, he had
arrived at the concept of a dynamical, four-dimensional Green function.  It
would be invaluable a century later.

To continue the saga of George Green and me---my next step was to trace the
influences of George Green on my own works.  Here I spent no time over ancient
documents.  I went directly to a known source:  THE WAR.

I presume that in Britain, unlike the United States, {\it the war} has a unique
connotation.  Apart from a brief sojourn in Chicago, to see if I wanted to help
develop The Bomb---I didn't---I spent the war years helping to develop
microwave radar.  In the earlier hands of the British, that activity, famous
for its role in winning the Battle of Britain, had begun with electromagnetic
radio waves of high frequency, to be followed by very high frequency, which led
to very high frequency, indeed.

Through those years in Cambridge (Massachusetts, that is), I gave a series of
lectures on microwave propagation.  A small percentage of them is preserved in
a slim volume entitled {\it Discontinuities in Waveguides}.  The word {\it
propagation} will have alerted you to the presence of George Green.  Indeed, on
pages 10 and 18 of an introduction there are applications of two different
forms of Green's identity.

Then, on the first page of Chapter 1, there is Green's function, symbolized by
G.  In the subsequent 138 pages the references to Green in name or symbol are
more than 200 in number.

As the war in Europe was winding down, the experts in high power microwaves
began to think of those electric fields as potential electron accelerators.  I
took a hand in that and devised the microtron which relies on the properties
of relativistic energy.  I have never seen one, but I have been told that it
works.  More important and more familiar is the synchrotron.

Here I was mainly interested in the properties of the radiation emitted by an
accelerated relativistic electron. I used the four-dimensionally invariant
proper time formulation of action.  It included the electromagnetic
self-action of the charge, which is to say that it employed a
four-dimensionally covariant Green's function.  I was only interested in the
resistive part, describing the flow of energy from the mechanical system into
radiation, but I could not help noticing that the mechanical mass had an
invariant electromagnetic mass added to it, thereby producing the physcial mass
of an electron.  I had always been told that such a union was not possible.
The simple lesson?  To arrive at covariant results, use a covariant
formulation, and maintain covariance throughout.

Quantum field theory, or more precisely, quantum electrodynammics, was forced
from childhood into adolescence by the experimental results announced at
Shelter Island early in June, 1947.  The relativistic theory of the electron
created by Dirac in 1928 was wrong.  Not very wrong, but measurably so.

A few days later, I left on a honeymoon tour across the United States.  Not
until September did I begin to work on the obvious hypothesis that
electrodynamic effects were responsible for the experimental deviations, one on
the magnetic moment of the electron, the other on the energy spectrum of the
hydrogen atom.

Although a covariant method was in order, I felt I could make up time with the
then more familiar non-covariant methods of the day.  By the end of November I
had the results.  The predicted shift in magnetic moment agreed with
experiment.  As for the energy shift in hydrogen, one ran into an expected
problem.

Consider the electromagnetic momentum associated with a charge moving at
constant speed.  The ratio of that momentum to the speed is a mass--an
electromagnetic mass.  It differs from the electromagnetic mass inferred from
the electromagnetic energy.  Analogously, the magnetic dipole moment inferred
for an electron moving in an electric field is wrong.  Replacing it by the
correct dipole moment leads to an energy level displacement that was correct
in 1947, and remains correct
today at that level of accuracy as governed by the fine structure constant.

I described all this at the January 1948 meeting of the American Physical
Society, after which Richard Feynman stood up and announced that he had a
relativistic method.  Well, so did I, but I also had the numbers.  Indeed,
several months later, at the opening of the Pocono Conference, he ran over to
me, shook my hand, and said ``Congratulations, Professor!  You got it right,"
which left me somewhat bewildered.  It turned out he had completed his own
calculation of the additional magnetic moment.  Later we compared notes and
found much in common.

Unfortunately, one of the things we shared was an incorrect treatment of low
energy photons.  Nothing fundamental was involved; it was a matter of technique
in making a transition between two different gauges.  But, as in American
politics these days, the less important the subject, the louder the noise.
When that lapse was set right, the result of 1947 was regained.  Incidentally,
even Lord Rayleigh once made a mistake.  That's one reason for its being called
the Rayleigh-Jeans law.

To keep to the main thrust of the talk---the evolution of Green's function in
the quantum mechanical realm---I move on to 1950, and a paper entitled {\it On
Gauge Invariance and Vacuum Polarization}.

This paper makes extensive use of Green's functions, in a proper-time context,
to deal with a variety of problems: non-linearities of the electromagnetic
field, the photon decay of a neutral meson, and a short, but not the shortest
derivation of the additional electron magnetic moment.  The latter ends with
the remark that ``The concepts employed here will be discussed at length in
later publications."  I cannot believe I wrote that.

The first, rather brief, discussion of those concepts appeared in a pair of
1951 papers, entitled {\it On the Green's Functions of Quantized Fields}.  One
would not be wrong to trace the origin of today's lecture back 42 years to
these brief notes.  This is how paper I begins:

The temporal development of quantized fields, in its particle aspect, is
described by propagation functions, or Green's functions.  The construction of
these functions for coupled fields is usually considered from the viewpoint of
perturbation theory.  Although the latter may be resorted to for detailed
calculations, it is desirable to avoid founding the formal theory of the
Green's functions on the restricted basis provided by the assumption of
expandability in powers of the coupling constants.  These notes are a
preliminary account of a general theory of Green's functions, in which the
defining property is taken to be the representation of the fields of prescribed
sources.

We employ a quantum dynammical principle for fields which has been described in
the 1951 paper entitled {\it The Theory of Quantized Fields}.  This (action)
principle is a differential characterization of the function that produces a
transformation from eigenvalues of a complete set of commuting operators on one
space-like surface to eigenvalues of another set on a different surface.

In one example of a rigorous formulation, Green's function, for an
electron-positron, obeys an inhomogeneous Dirac differential equation for an
electromagnetic vector potential that is supplemented by a functional
derivative with respect to the photon source; and, the vector potential obeys a
differential equation in which the photon source is supplemented by a vectorial
part of the electron-positron Green's function.  (It looks better than it
sounds.)  It is remarked that, in addition to such one-particle Green's
functions, one can also have multiparticle Green's functions.

The second note begins with:

In all the work of the preceding note there has been no explicit reference to
the particular states on (the space-like surfaces) that enter the definitions
of the Green's functions.  This information must be contained in boundary
conditions that supplement the differential equations.  We shall determine
these boundary conditions for the Green's functiions associated with vacuum
states on both (surfaces).

And then:

We thus encounter Green's functions that obey the temporal analog of the
boundary condition characteristic of a source radiating into space.  In keeping
with this analogy, such Green's functions can be derived from a retarded proper
time Green's function by a Fourier decomposition with respect to the mass.

The text continues with the introduction of auxiliary quantities: the mass
operator $M$ that gives a non-local extension to the electron mass; a somewhat
analogous photon polarization operator $P$; and $\Gamma$, the non-local
extension of the coupling between the electromagnetic field and the fields of
the charged particles.  Then, in the context of two-particle Green's functions,
there is the interaction operator $I$.

The various operators that enter in the Green's function equations $M$, $P$,
$\Gamma$, $I$, can be constructed by successive approximation.  Perturbation
theory, as applied in this manner, must not be confused with the expansion of
the Green's functions in powers of the charge.  The latter procedure is
restricted to the treatment of scattering problems.

Then one reads:

It is necessary to recognize, however, that the mass operator, for example, can
be largely represented in its effect by an alteration in the mass constant and
by a scale change of the Green's functiion.  Similarly, the major effect of the
polarization operator is to multiply the photon Green's function by a factor,
which everywhere appears associated with the charge.  It is only after these
renormalizations have been performed that we deal with wave equations that
involve the empirical mass and charge, and are thus of immediate physcial
applicability.

In the period 1951-1952, two colleagues of mine at Harvard, and I, wrote a
series of papers under the title {\it Electrodynamic Displacements of Atommic
Energy Levels}.  The third paper, which does not carry my name, is subtitled
{\it The Hyperfine Structure of Positronium}.  I quote a few lines:

The discussion of the bound states of the electron-positron system is based
upon a rigorous functional differential equation for the Green's function of
that system.

And,

Theory and experiment are in agreement.

As for the rest of the 50's, I focus on two highlights.  First: although it
could have appeared any time after 1951, it was 1958 when I published {\it The
Euclidean Structure of Relativistic Field Theory}.  Here is how it begins:

The nature of physcial experience is largely conditioned by the topology of
space-time, with its indefinite Lorentz metric.  It is somewhat remarkable,
then, to find that a detailed correspondence can be established between
relativistic quantum field theory and a mathematical image based on a
four-dimensional Euclidean manifold.  The objects that convey this
correspondence are the Green's functions of quantum field theory, which contain
all possible physcial information.  The Green's functions can be defined as
vacuum-state expectation values of time-ordered field products.

I well recall the reception this received, running the gamut from ``It's wrong"
to ``It's trivial."  It is neither.

Second (high light):

Another Harvard colleague and I had spent quite some time evolving the
techniques before we published a 1959 paper entitled {\it Theory of
Many-Particle Systems}.  It was intended to bring the full power of quantum
field theory to bear on the problems encountered in solid state physics, for
example.  That required the extension of vacuum Green's functions, which refer
to absolute zero temperature, into those for finite temperature.  This is
accomplished by a change of boundary conditions, which become statements of
periodicity, or anti-periodicity, for the respective BE or FD statistics, in
response to an imaginary time displacement.

As an off shoot of this paper, I published in 1960, {\it Field Theory of
Unstable Particles}.  Here is how it begins:

Some attention has been directed recently to the field theoretic description of
unstable particles.  Since this question is conceived as a basic problem for
field theory, the responses have been some special device or definition, which
need not do justice to the physical situation.  If, however, one regards the
descriptiion of unstable particles to be fully contained in the framework of
the general theory of Green's function, it is only necessary to emphasize the
relevant structure of these functions.  That is the purpose of this note.  What
is essentially the same question, the propagation of excitations in
many-particle systems where stable or long-lived ``particles" can occur under
exceptional circumstances, has already been discussed along thse lines.

One might be forgiven for assuming that this saga of George and me efectively
ended with this paper.  But that was 1/3 century ago!

To set the stage for what actually happened, I remind you that operator field
theory is an extrapolatiion of ordinary quantum mechanics, with its finite
number of degrees of freedom, to a continuum labeled by the spatial
coordinates.  The use of such space-time dependent variables presumes the
availability, in principle, of unlimited amounts of momentum and energy.  It
is,
therefore, a hypothesis about all possible phenomena of that type, the vast
majority of which lies far outside the realm of accessible physics.  In honor
of a failed economic policy, I call such procedures: trickle-down theory.

In the real world of physics, progress comes from tentative excursions beyond
the established framework of experiment and theory--the grass roots--indeed,
the Green grass roots.  What is sought here, in contrast with the speculative
approach of trickle-down theory, is a phenomenological theory--a coherent
account of the phenomena that is anabatic (from anabassis: going up).
\ve

The challenge was to reconstruct quantum field theory, without operator fields.
The source concept was introduced in 1951 as a mathematical device--it was a
source of fields.  It took 15 years to appreciate that, with a finite, rather
than an unlimited, supply of energy available, it made better sense to use the
more physical--if idealized--concept of a particle source.  Indeed, during that
time period one had become accustomed to the fact that to study a particle of
high energy physics, one had to create it.  And, the act of detection involved
the annihilation of that particle.

This idea first appeared in an article, entitled {\it Particles and Sources},
which recorded a lecture of the  1966 {\it Tokyo Summer Lectures in Theoretical
Physics}.  The preface begins with:

It is proposed that the phenomenological theory of particles be based on the
source concept, which is abstracted from the physical possibility of creating
or annihilating  any particle in a suitable collision.  The source
representation displays both the momentum (energy) and the space-time
characteristics of particle behavior.

Then, in the introduction, one reads:

Any particle can be created in a collision, given suitable partners, before and
after the impact to supply the appropriate values of the spin and other quantum
numbers, together with enough energy to exceed the mass threshold.  In
identifying new particles it is basic experimental principle that the specific
reaction is not otherwise relevant.  Then, let us abstract from the physical
presence of the additional particles involved in creating a given one (this is
the vacuum) and consider them simply as the source of the physcial properties
that are carried by the created particle.  The ability to give some
localization in space and time to a creation act may be represented by a
corresponding coordinate dependence of a mathematical source function, $S(x)$.
The effectiveness of the source in supplying energy and momentum may be
described by another mathematical source function, $S(p)$.  The complementarity
of these source aspects can be given its customary quantum interpretation:
$S(p)$ is the 4-dimensional Fourier transformation of $S(x)$.

The basic physical act begins with the creation of a particle by a source,
followed by the propagation (aha!) of that particle between the neighborhoods
of emission and detection, and is closed by the source annihilation of the
particle.  Relativistic requirements largely constrain the structure of the
propagation function---Green's function.

We now have a situation in which Green's function is not a secondary quantity,
implied by a more fundamental aspect of the theory, but rather, is a primary
part of the foundation of that theory.  Of course fields, initially inferred as
derivative concepts, are of great importance, as witnessed by the title I gave
to the set of books I began to write in 1968:  {\it Particles, Sources, and
Fields}.

The quantum electrodynamics that began to emerge in 1947 still bothers some
people because of the divergences that appear prior to renormalization.  That
objection is removed in the phenomenological source theory where there are no
divergences, and no renormalization.

As another example of such clarification I cite a 1975 paper entitled {\it
Casimir Effect in Source Theory}.  The abstract reads:

The theory of the Casimir effect, including its temperature dependence is
rederived by source theory methods, which do not employ the concept of
(divergent) zero point energy.  What source theory does have is a photon
Green's function, which changes in response to the change of boundary
conditions, as one conducting sheet is pushed into the proximity of another
one.

A few years later, I, and two colleagues at the University of California
(UCLA), who had joined me from Harvard with their new doctorates, extended this
treatment to dielectric bodies where forces of attraction also appear.

Having said this, I can move up to the present day, and the fascinating
phenomenon of coherent sonoluminescense.

It has only recently been discovered that a single air bubble in water can be
stabilized by an acoustical field.  And, that the bubble emmits pulses of
light,
including ultra violet light, in synchronism with the sonic frequency.

During the phase of negative acoustical pressure the bubble expands.  That is
followed by a contraction which, as Lord Rayleigh already recognized in his
1917 study of cavitation, turns into run away collapse.  The recent
measurements find speeds in excess of Mach 1 in air.

Then the collapse abruptly slows, and a blast of photons is emitted.  In due
time, the expansion slowly begins, and it all repeats, and repeats.

When confronted with a new phenomena, everyone tends to see in it something
that is already familiar.  So, when told about this new aspect of
sonoluminescence, I immediately said ``It's the Casimir effect!"  Not the
static Casimir effect, of course, but the dynamical one of accelerated
dielectric bodies.  I have had no occasion to change my mind.

I can imagine a member of this audience thinking: ``That's nice, but what is
the role of George Green in this?"

Looking in at the center of the water container, one sees a steady blue light.
A photomultiplier tube registers the succession of pulses, each containing a
substantial number of photons, which can be an incomplete count because, deep
in the ultraviolet, water becomes opaque.

A quantum mechanical description seeks the probabilities of emmitting various
numbers of photons, all of which probabilities are referred to the basic
probability, that for emitting no photons.  The latter probability dips below
one---in some analogy with synchrotron radiation---because of the self-action
carried by the electromagnetic field, as described by Green's functiion.  And
that function must obey the requirements imposed by an accelerated surface
discontinuity, with water, the dielectric material, on one side, and a
dielectric vacuum, air, on the other side.  Carrying out that program is---as
one television advertiser puts it--job one.  Very fascinating, indeed.

So ends our rapid journey through 200 years.  What, finally, shall we say about
George Green?  Why, that he is, in a manner of speaking, alive, well, and
living among us.

\bye